\documentclass[twocolumn,showpacs,amsmath,pra,aps,amssymb,longbibliography,nofootinbib]{revtex4-1}
\usepackage{graphicx,braket,float}
\usepackage{amsmath}
\usepackage{amsfonts}
\usepackage{amsbsy}
\usepackage{xcolor}
\usepackage{soul}
\usepackage{bm}
\usepackage[normalem]{ulem}
\usepackage[unicode]{hyperref}
\hypersetup{
    unicode=true, 
    plainpages=false,
    colorlinks=true,
    linkcolor=blue,
    citecolor=blue,
    filecolor=blue,
    urlcolor=blue
}
\newcommand{\cbEL}{\boldsymbol{\mathbf{\cal E}}}

\begin{document}
\title{Optical response of atom chains beyond the limit of low light intensity:\\ The validity of the linear classical oscillator model}
\author{L. A. Williamson}
\affiliation{Physics Department, Lancaster University, Lancaster LA1 4YB, UK} 
\author{J. Ruostekoski}
\affiliation{Physics Department, Lancaster University, Lancaster LA1 4YB, UK} 
\date{\today}

\begin{abstract}
Atoms subject to weak coherent incident light can be treated as coupled classical linear oscillators, supporting subradiant and superradiant collective excitation eigenmodes. We identify the limits of validity of this \emph{linear classical oscillator model} at increasing intensities of the drive by solving the quantum many-body master equation for coherent and incoherent scattering from a chain of trapped atoms. We show that deviations from the linear classical oscillator model depend sensitively on the resonance linewidths $\upsilon_\alpha$ of the collective eigenmodes excited by light, with the intensity at which substantial deviation occurs scaling as a powerlaw of $\upsilon_\alpha$. The linear classical oscillator model then becomes inaccurate at much lower intensities for subradiant collective excitations than superradiant ones, with an example system of seven atoms resulting in critical incident light intensities differing by a factor of 30 between the two cases. By individually exciting eigenmodes we find that this critical intensity has a $\upsilon_\alpha^{2.5}$ scaling for narrower resonances and more strongly interacting systems, while it approaches a $\upsilon_\alpha^3$ scaling for broader resonances and when the dipole-dipole interactions are reduced. The $\upsilon_\alpha^3$ scaling also corresponds to the semiclassical result whereby quantum fluctuations between the atoms have been neglected. We study both the case of perfectly mode-matched drives and the case of standing wave drives, with significant differences between the two cases appearing only at very subradiant modes and positions of Fano resonances.
\end{abstract}

\maketitle

\section{Introduction}
Light incident on closely spaced resonators (atoms, metamolecules, quantum dots, etc.) can scatter coherently multiple times between the resonators, resulting in strong light-mediated, long-range interactions. This provides a highly controllable system to explore classical and quantum many-body physics, which has been shown to boast subradiance~\cite{Guerin_subr16,Weiss19,Jenkins17,CAIT,bettles2015,Facchinetti16,Ritsch_subr,Jen17,Sutherland1D,Bettles2016a,Bhatti18,Asenjo_prx,Zhang2018,Plankensteiner2017,rui2020} and other collective phenomena in trapped atomic ensembles~\cite{BalikEtAl2013,Pellegrino2014a,Javanainen2014a,Jenkins2012a,CHA14,Kuraptsev14,dalibardexp,Yoo2016,Havey_jmo14,wilkowski,wilkowski2,Jenkins_thermshift,Dalibard_slab,vdStraten16,Machluf2018,antoine_polaritonic,Bettles18,Plankensteiner19,Javanainen19,Jen_2019,Kwong19,Binninger19} and in resonator arrays~\cite{Sentenac2008,lemoult2010,Trepanier17,Jenkins18}, with potential applications to quantum information processing~\cite{Grankin18,Guimond2019,ballantine2019}, the studies of nontrival topological phases~\cite{Bettles_topo,Perczel,perczel2018b}, and atomic clocks~\cite{Ye2016,Kramer2016,Henriet2018,Qu19}.

The exponentially large Hilbert space of a quantum many-atom system has confined most theoretical studies of collective optical response to the low light intensity regime. The dynamics of two-level atoms then reduces to that of a collection of coupled linear classical oscillators~\cite{Javanainen1999a,Lee16}, or a model of coupled dipoles. In this regime the polarizations evolve linearly as a function of a coherent drive field, and can therefore be understood by dividing the system into collective eigenmodes $\mathbf{u}_\alpha$, $\alpha=1,...,N$, with corresponding radiative resonance linewidths $\upsilon_\alpha$. For a dense ensemble of cold atoms, strong light-induced correlations between different atoms  in the limit of low light intensity emerge from the fluctuations of atomic positions~\cite{Morice1995a,Ruostekoski1997a}, while for the atoms at fixed positions the correlations are absent.

Solving the full quantum dynamics beyond the limit of low light intensity is computationally much more demanding, and hence many-body quantum effects on scattering has seen little exploration. Examples of full quantum studies include the spectra of small arrays of atoms~\cite{Olmos16} and the identification of quantum effects in the transmission of light through planar arrays of atoms~\cite{bettles2019}. 
Importantly, the precise limits of validity of the linear classical oscillator model and the onset of quantum fluctuations in the low-excitation regime have not been addressed in strongly coupled many-atom systems. Experiments are frequently modeled using the linear classical oscillator model, even though reaching the weak excitation limit in small atomic ensembles may sometimes in practice be challenging. Therefore determining the limits of validity of the approximation is relevant for interpreting experimental findings.

In this work we explore light scattering from one-dimensional atomic chains~\cite{Sutherland1D,Bettles2016a,Asenjo_prx,Zhang2018,Needham19,Cardoner19,Clemens2003a,kien2008,kien2014,ruostekoski2017,albrecht2019,jones2019,Zhang19} of subwavelength spaced two-level atoms via simulations of the full quantum many-body master equation. Collective optical responses of regular arrays of atoms have now been experimentally measured for the case of 
a planar optical lattice in a Mott-insulator state, demonstrating subradiant resonance narrowing~\cite{rui2020}.
By simulating both coherent and incoherent scattering for the atomic chain as a function of drive strength, we identify the regimes of validity of the linear classical oscillator model. We find that the low light intensity collective linewidths $\upsilon_\alpha$ play a crucial role in determining when the coherent scattering deviates from that of radiating linear oscillators. In particular, the critical intensity $\mathcal{I}_C$ at which this deviation becomes appreciable is much smaller for drive fields mode-matched and resonant with subradiant collective modes than superradiant ones. Even in small example systems ($\lesssim 10$ atoms), $\mathcal{I}_C$ can dramatically differ by up to two orders of magnitude between the most subradiant and superradiant modes, imposing very different conditions for the validity of simulations using linear classical oscillator models.

We simulate the optical response from a variety of lattice spacings, orientations and atom numbers, and find that $\mathcal{I}_C$ scales as a powerlaw of $\upsilon_\alpha$. We firstly consider drive fields exactly mode-matched to the low light intensity collective modes $\mathbf{u}_\alpha$. For modes with narrow resonances $\upsilon_\alpha\lesssim 0.5\gamma$ (with $\gamma$ the single atom linewidth), or all modes in the limit of strong light-mediated interactions (small array spacing), we find that $\mathcal{I}_C\propto\upsilon_\alpha^{2.5}$. The remaining modes scale as $\mathcal{I}_C\propto\upsilon_\alpha^3$. Semiclassical simulations, whereby quantum fluctuations between the atoms have been neglected, give $\mathcal{I}_C\propto\upsilon_\alpha^3$, for all the cases, which reproduces the superradiant scaling outside the regime of very strong interactions. We find that the critical intensity $\mathcal{I}_I$ at which incoherent scattering becomes appreciable follows closely the behavior of $\mathcal{I}_C$. We extend our analysis to standing wave drive fields, with the incident angle of the drive chosen to maximize overlap with each of the low light intensity collective modes. For modes with wavelength inside the light line, the standing wave drive gives quantitatively similar results to the perfectly mode-matched case, except at positions of Fano resonances.

The paper is organized as follows. In Sec.~\ref{background} we introduce the system setup and provide necessary background details. In Sec.~\ref{lowlightbreak} we study the validity of the linear classical oscillator model for drive fields mode-matched and resonant with different low light intensity collective modes. We study the deviation between the coherent scattering predicted by the linear classical oscillator model and the full quantum solution, and identify the intensity at which incoherent scattering becomes appreciable. In Sec.~\ref{realisticDrives} we compare the previous results with those obtained using standing wave drives. We conclude in Sec.~\ref{conclusion}.

\section{Background}\label{background}
\subsection{System setup}
\begin{figure}
\includegraphics[trim=9cm 2cm 7cm 5cm,clip=true,width=0.48\textwidth]{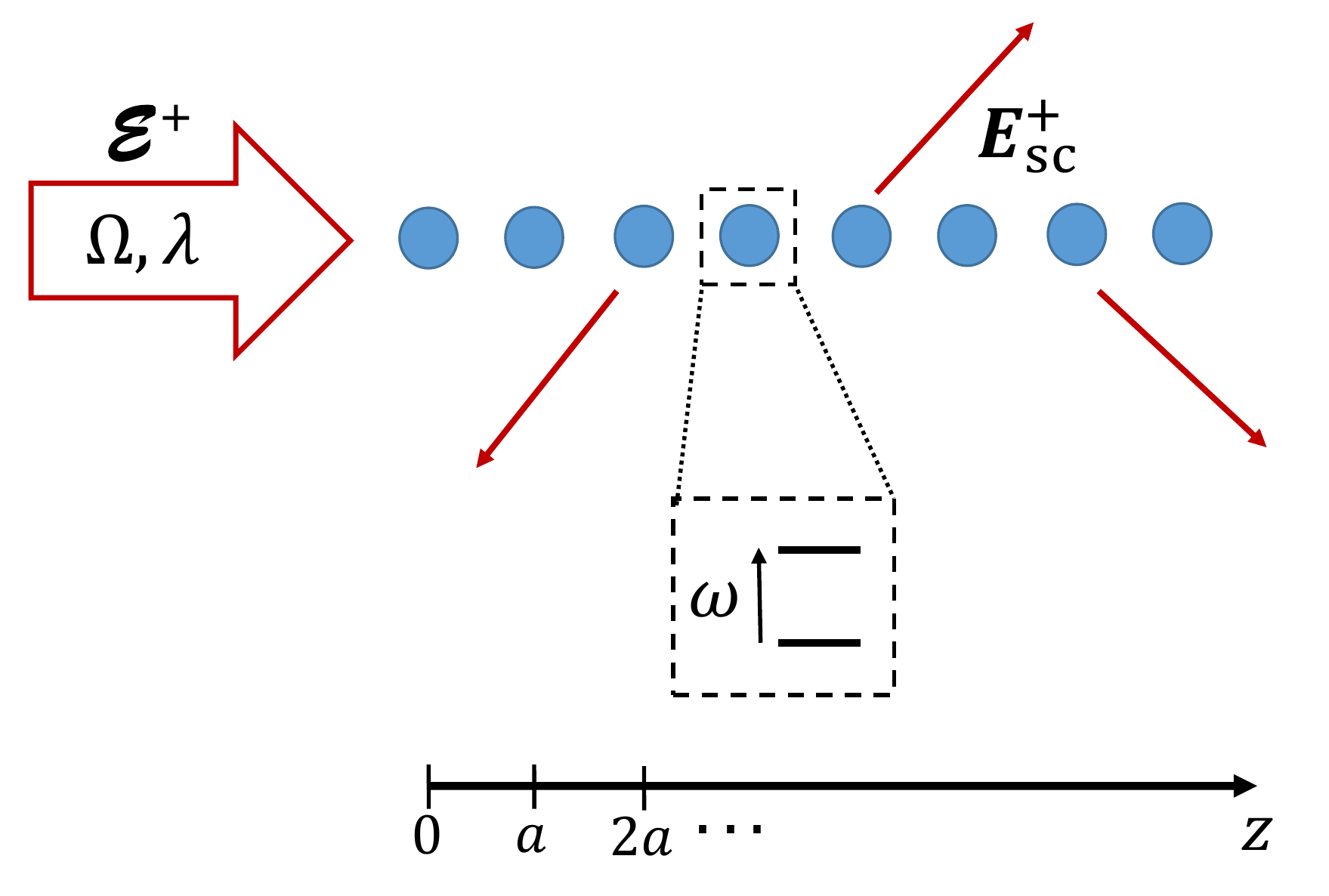}
\caption{\label{system}System setup. A coherent field is incident on a chain of two-level atoms oriented along $\hat{\mathbf{z}}$ with spacing $a<\lambda$. A detector completely enclosing the atoms collects all of the scattered light $\mathbf{E}_s^+$, to give a photon scattering rate $n$. Analyzing the coherent $n_C$ and incoherent $n_I$ photon scattering rates allows a systematic study of the limits of validity of the linear classical oscillator model, as a function of incident intensity $I_\text{in}$.}
\end{figure}

We consider the dynamics of a regular chain of $N$ identical two-level atoms with positions $\mathbf{r}_m=a(m-1)\hat{\mathbf{z}}$, $m=1,...,N$, driven by a coherent drive $\cbEL^+(\mathbf{r})e^{i\Omega t}$, see Fig.~\ref{system}. The positive frequency component of the light amplitude $\cbEL^+(\mathbf{r}) ={\bf D}_F^+(\mathbf{r})/\epsilon_0$ (where ${\bf D}_F^+(\mathbf{r})$ denotes the electric displacement outside the atoms) along the direction $\hat{\mathbf{e}}$ of the dipole moments  of the atoms is parameterized as
\begin{align}
\cbEL^+(\mathbf{r})=\frac{1}{2}\mathcal{E}_0\psi(\mathbf{r})\hat{\mathbf{e}}
\end{align}
with the spatial dependence $\psi(\mathbf{r})$ normalized as $\sum_m |\psi(\mathbf{r}_m)|^2=1$ and $\cbEL^-=(\cbEL^+)^*$. The incident field intensity averaged over the atoms is
\begin{align}
I_\text{in}=\frac{1}{N}\sum_m 2\epsilon_0 c|\cbEL^+(\mathbf{r}_m)|^2=\frac{\epsilon_0 c|\mathcal{E}_0|^2}{2N}.
\end{align}

The dynamics of the atoms in a frame rotating at the driving field frequency $\Omega$ is described, in the length gauge~\cite{PowerZienauPTRS1959,woolley1971,CohenT,Ruostekoski1997a}, by the master equation for the reduced density matrix $\rho$~\cite{Lehmberg1970,agarwal1970},
\begin{align}\label{masterEq}
\frac{d\rho}{dt}&=-\frac{i}{\hbar}\sum_m[H_m,\rho]-i\sum_m\sum_{n\ne m}\Delta_{mn}[\sigma_m^+\sigma_n^-,\rho]\nonumber\\
&\phantom{=}+\sum_{m,n}\mathcal{L}_{mn}[\rho],
\end{align}
with
\begin{align}
H_m&=\hbar\left[\delta\sigma_m^{ee}+\mathcal{R}^*\psi^*(\mathbf{r}_m)\sigma_m^-+\mathcal{R}\psi(\mathbf{r}_m)\sigma_m^+\right],\nonumber\\
\mathcal{L}_{mn}[\rho]&=\gamma_{mn}\left(2\sigma_n^-\rho\sigma_m^+-\sigma_m^+\sigma_n^-\rho-\rho\sigma_m^+\sigma_n^-\right).
\end{align}

Here and below all summation indices run over all $N$ atoms, unless otherwise indicated. The Hamiltonian $H_m$ and single atom decay terms $\mathcal{L}_{mm}$ describe the single atom dynamics, while the remaining terms described the light-mediated interactions. For each atom $m$, $\sigma_m^+$ and $\sigma_m^-$ are spin $1/2$ raising and lowering operators, respectively, $\sigma_m^{ee}=\sigma_m^+\sigma_m^-$ is the excited state population operator, and $\mathcal{R}\psi(\mathbf{r}_m)=\psi(\mathbf{r}_m)\mathcal{D}\mathcal{E}_0/(2\hbar)$ is the complex Rabi frequency, with $\mathcal{D}$ the reduced dipole matrix element that without loss of generality we choose to be real. We have made the rotating wave approximation in $H_m$ by omitting the fast co-rotating terms $\mathcal{R}^*\psi^*(\mathbf{r}_m)\sigma_m^- e^{2i\Omega t}+\mathcal{R}\psi(\mathbf{r}_m)\sigma_m^+ e^{-2i\Omega t}$.  The two-level transition frequency $\omega$ is detuned from the drive field frequency by $\delta=\omega-\Omega$, and $\gamma_{mm}=\gamma=\mathcal{D}^2k^3/(6\pi\hbar\epsilon_0)$, where $k=2\pi/\lambda$ is the resonant wavenumber of the incident light, with $c$ the speed of light in vacuum and $\lambda=c/\omega$ the resonant wavelength. 
We neglect recoil effects from the light scattering (which in dense atom clouds can also be correlated~\cite{Robicheaux19})  and have considered the atoms as being stationary. We also assume the lattice confinement is sufficiently tight that position fluctuations of the atoms can be ignored.

The light-mediated coherent and dissipative interactions are obtained from the dipolar scattering kernel $\mathsf{G}(\mathbf{r})$, with respective strengths
\begin{align}\label{Deltadef}
\Delta_{mn}&=\frac{1}{\hbar\epsilon_0}\operatorname{Re}\left[\mathbf{d}^*\cdot \mathsf{G}(\mathbf{r}_{mn})\mathbf{d}\right],
\end{align}
\begin{align}\label{Gammadef}
\gamma_{mn}&=\frac{1}{\hbar\epsilon_0}\operatorname{Im}\left[\mathbf{d}^*\cdot \mathsf{G}(\mathbf{r}_{mn})\mathbf{d}\right],
\end{align}
with $\mathbf{r}_{mn}=\mathbf{r}_m-\mathbf{r}_n$, $\mathbf{d}=\mathcal{D}\hat{\mathbf{e}}$ and~\cite{Jackson}
\begin{align}\label{Gdef}
\mathsf{G}(\mathbf{r})\mathbf{d}&=-\frac{\mathbf{d}\delta(\mathbf{r})}{3}+\frac{k^3}{4\pi}\Bigg\{\left(\hat{\mathbf{r}}\times\mathbf{d}\right)\times\hat{\mathbf{r}}\frac{e^{ikr}}{kr}\nonumber\\
&\phantom{==}-\left[3\hat{\mathbf{r}}\left(\hat{\mathbf{r}}\cdot\mathbf{d}\right)-\mathbf{d}\right]\left[\frac{i}{(kr)^2}-\frac{1}{(kr)^3}\right]e^{ikr}\Bigg\}.
\end{align}
(with $\hat{\mathbf{r}}=\mathbf{r}/|\mathbf{r}|$). Note that $\gamma_{mm}=\gamma$ whereas $\Delta_{mm}$ diverges. The divergence of $\Delta_{mm}$ can be accounted for by a proper treatment of the Lamb shift, which we assume has been absorbed into the single atom detuning $\delta$. Light-mediated interactions are significant when the lattice spacing $a$ satisfies $a\lesssim \lambda$. Due to the anisotropic radiation profile of an oscillating dipole, the orientation and polarization of the incident drive relative to the lattice direction $\hat{\mathbf{z}}$ affects the strength of the interactions. We take $\hat{\mathbf{e}}$ to be a circularly polarized unit vector. For light incident parallel to the atom chain we have $\hat{\mathbf{e}}=(\hat{\mathbf{x}}-i\hat{\mathbf{y}})/\sqrt{2}$ and so $\hat{\mathbf{z}}\cdot\hat{\mathbf{e}}=0$. For light incident perpendicular to the atom chain we have $\hat{\mathbf{e}}=(\hat{\mathbf{z}}-i\hat{\mathbf{x}})/\sqrt{2}$ and $|\hat{\mathbf{z}}\cdot\hat{\mathbf{e}}|=1/\sqrt{2}$.

\subsection{Scattered light properties}

The total field outside of the atoms is the sum of the incident and scattered fields,
\begin{align}
\mathbf{E}^\pm(\mathbf{r})=\cbEL^\pm(\mathbf{r})+\mathbf{E}_\text{sc}^\pm(\mathbf{r}),
\end{align}
with
\begin{align}\label{Escat}
\epsilon_0\mathbf{E}_\text{sc}^\pm({\textbf{r}})=\sum_m \mathsf{G}(\mathbf{r}-\mathbf{r}_m)\mathbf{d}\sigma_m^\mp.
\end{align}
The scattered field consists of both a mean field $\left<\mathbf{E}_\text{sc}^\pm\right>$ and fluctuations $\delta\mathbf{E}_\text{sc}^\pm=\mathbf{E}_\text{sc}^\pm-\left<\mathbf{E}_\text{sc}^\pm\right>$. The total light intensity outside of the atoms is
\begin{align}
I(\mathbf{r})=2\epsilon_0 c\left<\mathbf{E}^-(\mathbf{r})\cdot\mathbf{E}^+(\mathbf{r})\right>,
\end{align}
with
\begin{align}\label{Ecor}
\left<\mathbf{E}^-(\mathbf{r})\cdot\mathbf{E}^+(\mathbf{r})\right>=&\cbEL^+(\mathbf{r})\cdot\cbEL^-(\mathbf{r})+\cbEL^+(\mathbf{r})\cdot\left<\mathbf{E}_\text{sc}^-(\mathbf{r})\right>\nonumber\\
&\left<\mathbf{E}_\text{sc}^+(\mathbf{r})\right>\cdot\cbEL^-(\mathbf{r})+\left<\mathbf{E}_\text{sc}^+(\mathbf{r})\right>\cdot\left<\mathbf{E}_\text{sc}^-(\mathbf{r})\right>\nonumber\\
&+\left<\delta\mathbf{E}_\text{sc}^+(\mathbf{r})\cdot\delta\mathbf{E}_\text{sc}^-(\mathbf{r})\right>.
\end{align}
The first term in Eq.~\eqref{Ecor} is the incident field contribution to the intensity. The next two terms give the interference between the incident field and the coherently scattered field, which can be used for a homodyne measurement of the coherent scattered field. The fourth term is proportional to the coherent scattered light intensity and the fifth term the incoherently scattered light intensity, which, for the case of fixed atomic positions, arises solely from quantum fluctuations.

We assume that the incident field has been blocked before its photons are detected, for example by a thin wire as in the dark-ground imaging technique~\cite{andrews1996}. Only the scattered light intensity $I_\text{sc}$ is therefore detected, with
\begin{align}
I_\text{sc}(\mathbf{r})=2\epsilon_0 c\left(\left<\mathbf{E}_\text{sc}^-(\mathbf{r})\right>\cdot\left<\mathbf{E}_\text{sc}^+(\mathbf{r})\right>+\left<\delta\mathbf{E}_\text{sc}^-(\mathbf{r})\cdot\delta\mathbf{E}_\text{sc}^+(\mathbf{r})\right>\right).
\label{intensity}
\end{align}
The photon count-rate integrated over a detector surface $S$ is then
\begin{align}\label{scatteringrate1}
n=\frac{1}{\hbar\omega}\int_S dS\,I_\text{sc}(\mathbf{r}),
\end{align}
where $dS$ denotes an area element. The photon count-rate is made up of two contributions: a coherent scattering rate,
\begin{align}\label{scatteringratec}
n_C=\frac{2\epsilon_0 c}{\hbar\omega}\int_S dS\,\left<\mathbf{E}_\text{sc}^-(\mathbf{r})\right>\cdot\left<\mathbf{E}_\text{sc}^+(\mathbf{r})\right>,
\end{align}
that survives in the absence of fluctuations $\mathbf{E}_\text{sc}^\pm\rightarrow\left<\mathbf{E}_\text{sc}^\pm\right>$; and an incoherent scattering rate,
\begin{align}\label{scatteringratei}
n_I=\frac{2\epsilon_0 c}{\hbar\omega}\int_S dS\,\left<\delta\mathbf{E}_\text{sc}^-(\mathbf{r})\cdot\delta\mathbf{E}_\text{sc}^+(\mathbf{r})\right>.
\end{align}

For simplicity we assume that the detector completely encloses the atoms, so that all the scattered light is collected. The surface integral in Eqs.~\eqref{scatteringrate1}--\eqref{scatteringratei} can then be evaluated analytically to give~\cite{carmichael2000} (see Appendix)
\begin{align}\label{scatteringrate}
n&=2\sum_{m,n}\gamma_{mn}\left<\sigma_m^+\sigma_n^-\right>,\nonumber\\
n_C&=2\sum_{m,n}\gamma_{mn}\left<\sigma_m^+\right>\left<\sigma_n^-\right>,\nonumber\\
n_I&=2\sum_{m,n}\gamma_{mn}\left(\left<\sigma_m^+\sigma_n^-\right>-\left<\sigma_m^+\right>\left<\sigma_n^-\right>\right).
\end{align}
Equation~\eqref{scatteringrate} also follows immediately from Eq.~\eqref{masterEq}, as the loss rate of excitations, which equals the photon detection rate $n$ (when all photons are detected), is $\sum_{m,n,k}\operatorname{Tr}(\sigma_k^+\sigma_k^-\mathcal{L}_{mn}[\rho])=2\sum_{m,n}\gamma_{mn}\left<\sigma_m^+\sigma_n^-\right>$.

For atoms with fixed positions, a nonzero incoherent scattering rate requires one of two things: either nonnegligible population in the excited levels of one or more atoms, $\left<\sigma_m^{ee}\right>>0$; or nonnegligible many-body correlations $\left<\sigma_m^+\sigma_n^-\right>\ne \left<\sigma_m^+\right>\left<\sigma_n^-\right>$ for $n\ne m$, which can be generated via the light-mediated interactions. For $N=1$ both the steady-state coherent and incoherent scattering rates can be solved analytically~\cite{Mollow1969a},
\begin{align}\label{nSingleAtom}
n_C&=\frac{\gamma I_\text{in}/I_s}{(1+\delta^2/\gamma^2+I_\text{in}/I_s)^2},\nonumber\\
n_I&=\gamma\left(\frac{I_\text{in}/I_s}{1+\delta^2/\gamma^2+I_\text{in}/I_s}\right)^2,
\end{align}
where
\begin{align}
I_s=\frac{\hbar ck^3\gamma}{6\pi}
\end{align}
is the single atom saturation intensity. Note $I_\text{in}/I_s=2|\mathcal{R}|^2/\gamma^2$, hence $I_\text{in}\sim I_s$ equates to driving the system such that the magnitude of the Rabi frequency is comparable to the single atom linewidth. Assuming a resonant drive, for low drive intensities $I_\text{in}\ll I_s$ we can expand Eq.~\eqref{nSingleAtom} in $I_\text{in}/I_s$ to give $n_C\propto I_\text{in}$ and $n_I\propto I_\text{in}^2$. In this linear regime the coherent scattering dominates, $n_C\gg n_I$. Nonlinear effects become notable for $I_\text{in}\gtrsim I_s$.

\subsection{Limit of low light intensity: the linear classical oscillator model}
In the limit of low light intensity we can neglect terms that contain two or more excited state field amplitudes or one or more excited state amplitudes multiplied by the driving field, where the field amplitudes refer to second quantized atomic fields~\cite{Ruostekoski1997a}. In our system, this amounts to neglecting terms $\left<\sigma_m^+\sigma_n^-\right>$ (along with higher order correlators of $\sigma^\pm_m$) and $\mathcal{R}\left<\sigma_m^+\right>$. The only nontrival elements of the density matrix are then $\rho_m^{ge}(t)=\left<\sigma_m^-(t)\right>$, and Eq.~\eqref{masterEq} reduces to,
\begin{align}\label{lowLight}
\frac{d\rho_m^{ge}}{dt}&=i\sum_k\mathcal{H}_{mk}\rho_k^{ge}+i\mathcal{R}\psi( \mathbf{r}_m).
\end{align}
The diagonal elements of the matrix $\mathcal{H}$ describe the single atom detuning and linewidth, $\mathcal{H}_{mm}=\delta+i\gamma$, while the off-diagonal elements arise from the low light intensity dipole-dipole interactions, $\mathcal{H}_{mk}= \Delta_{mk}+i\gamma_{mk}$, for $k\ne m$. Equation~\eqref{lowLight} is identical to the equation for classical coupled dipoles.

Consistently, the limit of  low light intensity for the optical response is obtained from Eq.~\eqref{Ecor} by calculating the coherently scattered light intensity to the lowest order in the field amplitude, 
\begin{align}\label{Ecor2}
\left<\mathbf{E}^-(\mathbf{r})\cdot\mathbf{E}^+(\mathbf{r})\right>=&\cbEL^+(\mathbf{r})\cdot\cbEL^-(\mathbf{r})+\cbEL^+(\mathbf{r})\cdot\left<\mathbf{E}_\text{sc}^-(\mathbf{r})\right>\nonumber\\
&\left<\mathbf{E}_\text{sc}^+(\mathbf{r})\right>\cdot\cbEL^-(\mathbf{r})+\mathcal{O}\left[\left|\left<\mathbf{E}_\text{sc}^+(\mathbf{r})\right>\right|^2\right].
\end{align}
The field amplitude $\left<\mathbf{E}_\text{sc}^+(\mathbf{r})\right>$ in Eq.~\eqref{Ecor2} is obtained from Eq.~\eqref{Escat} after solving for $\left<\sigma_m^-\right>$ using Eq.~\eqref{lowLight}.\footnote{Equation~\eqref{lowLight} gives the dynamics of atoms at fixed positions $\mathbf{r}_1,\mathbf{r}_2,\ldots, \mathbf{r}_N$. One can formally show that the model provides an exact solution for coherently scattered light of laser-driven atoms also for the case of stochastically distributed atomic positions in the limit of low light intensity~\cite{Javanainen1999a,Lee16}. In that case, the linear classical oscillator model is solved for each stochastic realization of fixed atomic positions $\mathbf{r}_1,\mathbf{r}_2,\ldots, \mathbf{r}_N$ that are sampled from the appropriate distribution, and the calculated optical response is then ensemble-averaged over many such runs.}

The complex symmetric matrix $\mathcal{H}$ affords a complete but not necessarily orthogonal basis of eigenstates $\{\mathbf{u}_\alpha\}$ ($\alpha=1,...,N$), which are the low light intensity collective excitation eigenmodes~\cite{JenkinsLongPRB}. We assume the $\mathbf{u}_\alpha$ vectors are normalized, $|\mathbf{u}_\alpha|=1$. The corresponding complex eigenvalues $\{\zeta_\alpha+i\upsilon_\alpha\}$ have real part $\zeta_\alpha$, which is the collective mode resonance shift from the resonance of an isolated atom, and imaginary part $\upsilon_\alpha$, which is the collective linewidth. Collective modes with broad resonances $\upsilon_\alpha>\gamma$ are termed superradiant, while those with narrow resonances $\upsilon_\alpha<\gamma$ are termed subradiant. The range of collective linewidths can span many orders of magnitude~\cite{Jenkins2012a,Asenjo_prx,Jenkins_thermshift,Sutherland1D,Bettles2016a}, hence the radiative properties of weakly excited atomic ensembles vary greatly depending on whether subradiant or superradiant modes are excited. If the field profile of the drive is chosen to match a collective mode, $\psi(\mathbf{r}_m)=u_{\alpha m}$, with $u_{\alpha m}$ the $m$th component of $\mathbf{u}_\alpha$, the steady-state polarization is simply,
\begin{align}
\left<\sigma_m^-\right>=\frac{\mathcal{R}}{\zeta_\alpha+i\upsilon_\alpha}u_{\alpha m}.
\end{align}
This is identical to the polarization of a single harmonic oscillator but with the single atom detuning and linewidth replaced by the collective mode detuning and linewidth.

Here we investigate the linear classical oscillator model~\eqref{lowLight} beyond the low light intensity regime \eqref{Ecor2} by studying the coherent scattering rate $n_C$, which is second order in $\cbEL^+(\mathbf{r})$ (or $\left<\mathbf{E}_\text{sc}^+(\mathbf{r})\right>$). The linear classical oscillator model predicts a linear dependence of $n_C$ on the incident intensity $I_\text{in}$. In the limit that the incoherent scattering is dominated by position fluctuations of the atoms, the linear classical oscillator model can also represent the incoherently scattered light intensity, in which case the $\left<\delta\mathbf{E}_\text{sc}^-(\mathbf{r})\cdot\delta\mathbf{E}_\text{sc}^+(\mathbf{r})\right>$ contribution is entirely generated by the fluctuating positions of the atoms. For atoms at fixed positions, as considered in this paper, the incoherent scattering from the linear classical oscillator model vanishes, and all fluctuations are solely due to quantum effects.

\section{Validity of the linear classical oscillator model}\label{lowlightbreak}

We investigate the limits of validity of the linear classical oscillator model by comparing its predictions for light scattering with predictions from the full quantum many-body master equation~\eqref{masterEq}, as a function of drive intensity, the atom number, and the atomic spacing. For simplicity, we examine the steady-state optical responses. Coherent scattering deviating from a linear dependence on $I_\text{in}$ signifies discrepancies from the linear classical oscillator model, in which case the model no longer provides an accurate description of the scattering. The presence of appreciable incoherent scattering also directly implies dynamics beyond the linear classical oscillator model. For a given atom number $N$, we drive the system with a field mode-matched and resonant with different collective modes $u_\alpha$, i.e.\ $\psi(\mathbf{r}_m)=u_{\alpha m}$, $\zeta_\alpha=0$, for $\alpha=1,...,N$. For the most subradiant modes in systems with a very small lattice spacing such a drive field is an idealization, as the phase variation required for the field is too rapid (subwavelength). We consider standing-wave drive fields in Sec.~\ref{realisticDrives}, where we will also be able to address how realistic fields will affect the excitation of such eigenmodes.

\subsection{Coherent scattering}

\begin{figure}

\includegraphics[trim=0cm 1cm 0cm 2cm,clip=true,width=0.45\textwidth]{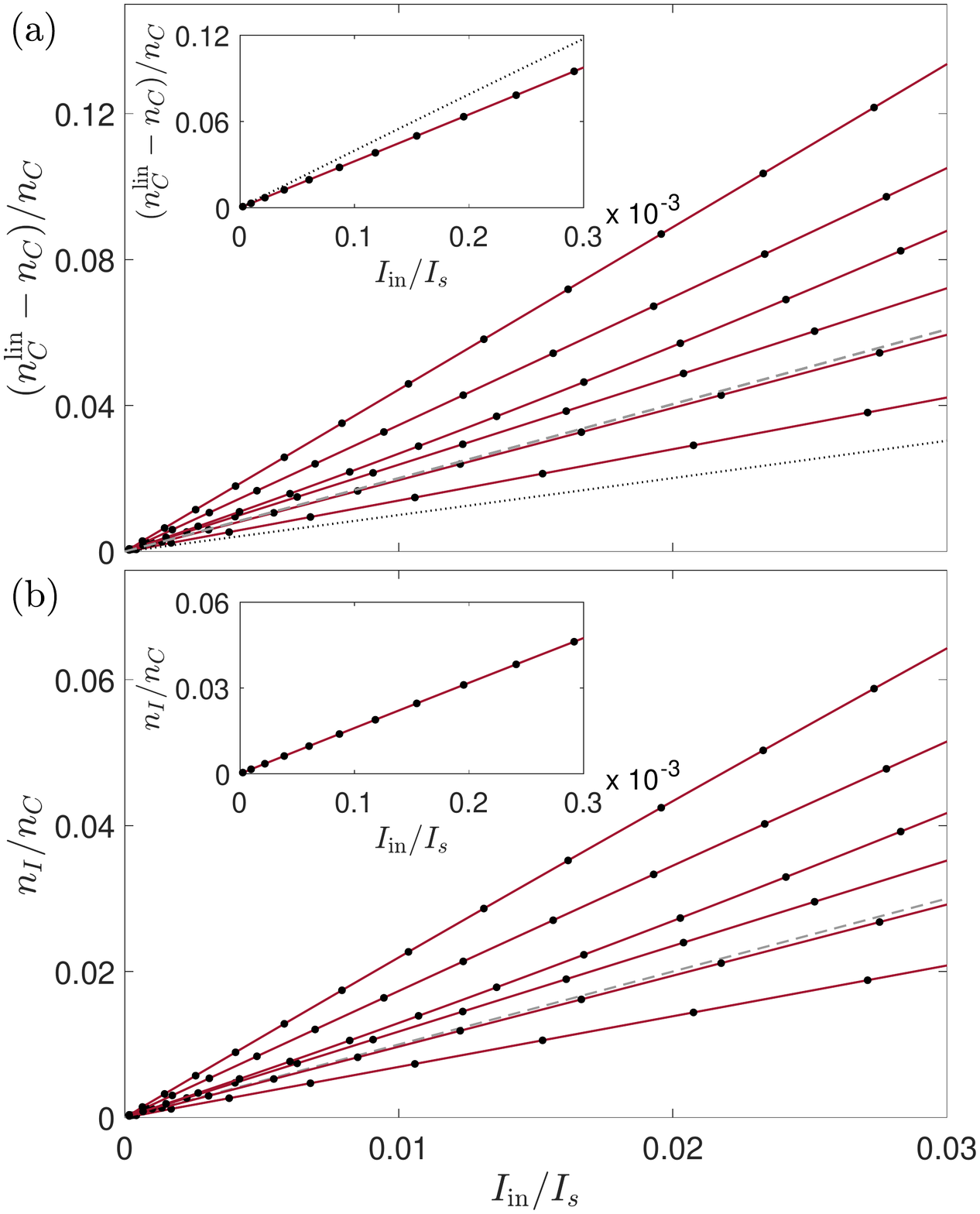}
\caption{\label{cohSca}Deviation between the full quantum scattering and scattering predicted from the linear classical oscillator model. Parameters are $N=7$, $a=0.4\lambda$ and $\hat{\mathbf{e}}=(\hat{\mathbf{x}}-i\hat{\mathbf{y}})/\sqrt{2}$. The separate curves correspond to drive fields that are mode-matched and resonant with different low light intensity collective modes $\mathbf{u}_\alpha$ (dots are numerical simulations, red curves are spline fits). In (a) and (b) the corresponding linewidths are, from shallowest curve to steepest curve, $\upsilon_\alpha=(1.41,1.29,1.15,0.97,1.03,0.96)\gamma$ (main figures) and $\upsilon_\alpha=0.18\gamma$ (insets). (a) The relative deviation between the coherent scattering rate obtained from the linear classical oscillator model ($n_C^\text{lin}$) and the full quantum solution ($n_C$) grows proportional to $I_\text{in}$.  In all but one case the slope of the curves depends inversely on the collective mode linewidth. The semiclassical model (dotted black curve) for a drive mode-matched and resonant with the most superradiant mode (main figure) and most subradiant mode (inset) captures the qualitative behavior of the scattering but is not quantitatively accurate. (b) The ratio of incoherent ($n_I$) to coherent scattering displays the same behavior as $(n_C^\text{lin}-n_C)/n_C$. The results for a single isolated atom, obtained from Eq.~\eqref{nSingleAtom}, are shown by the gray dashed lines in (a) and (b) for comparison.}
\end{figure}

In Fig.~\ref{cohSca}(a) we plot the relative deviation between the coherent scattering rate $n_C$, obtained from the steady-state solution of the full quantum description~\eqref{masterEq}, and the coherent scattering rate $n_C^\text{lin}$, obtained from the steady-state solution of the linear classical oscillator model~\eqref{lowLight}, as a function of drive intensity, for $N=7$ atoms. We do this for drives mode-matched and resonant with each of the $N$ different low light intensity collective modes. The relative deviation $(n_C^\text{lin}-n_C)/n_C$ clearly scales proportional to $I_\text{in}$ for the intensities shown. We find that the slope $(n_C^\text{lin}-n_C)/(n_C I_\text{in})$ has a pronounced dependence on which collective mode is driven, with a smaller $\upsilon_\alpha$ resulting in a larger slope in all but one case.\footnote{The exception occurs at the $\upsilon_\alpha=0.97\gamma$ curve, which is less steep than the $\upsilon_\alpha=1.03\gamma$ curve.} The drive mode-matched to the $\upsilon_\alpha=0.96\gamma$ mode, for example, results in appreciable deviation between $n_C$ and $n_C^\text{lin}$ at intensities $I_\text{in}\gtrsim 0.02I_s$, whereas a drive mode-matched to the most superradiant mode ($\upsilon_\alpha=1.41\gamma$) shows similar deviation at thrice this intensity. A more extreme case occurs when the lattice spacing is increased to $a=0.6\lambda$. Then, driving the subradiant mode with $\upsilon_\alpha=0.63\gamma$ results in substantial deviation between $n_C$ and $n_C^\text{lin}$ at intensities $I_\text{in}\gtrsim 0.01I_s$, whereas driving the superradiant mode with $\upsilon_\alpha=2.0\gamma$ gives similar deviation only at much higher intensities $I_\text{in}\gtrsim 0.3I_s$. The low light intensity collective modes therefore play a crucial role in determining the response of the atoms not just within the regime of validity of the linear classical oscillator model, but also when quantum and nonlinear effects start becoming important due to increasing light intensity.\footnote{Note that perfect mode matching to the most subradiant mode in Fig.~\ref{cohSca} is an idealization, as the required phase variation is too rapid. We will later show how this is modified when a realistic drive is used instead.}

For stochastically distributed atoms in ensembles where light induces strong spatial correlations between the atoms (due to their fluctuating positions), a semiclassical (SC) model where all quantum fluctuations between the atoms are neglected~\cite{Lee16}  has provided in several regimes of interest a numerically tractable description for the full quantum dynamics of the system~\cite{bettles2019,Machluf2018,Sutherland_satur,Lee17}, as well as a systematic mechanism to unambiguously identify quantum effects in the scattered light~\cite{bettles2019}. For atoms at fixed positions, as we are interested in here, the analogous approximation corresponds to neglecting all correlations between the atoms and obtaining a mean-field-theoretical solution~\cite{Kramer2015a,Parmee2018}. This is obtained by explicitly factorizing all correlations between atoms in the equations of motion~\eqref{masterEq}, $\left<\sigma_m^\mu\sigma_n^\nu\right>\rightarrow \left<\sigma_m^\mu\right>\left<\sigma_n^\nu\right>$ for $\mu,\nu\in\{+,-,ee\}$ and $n\ne m$. Equation~\eqref{masterEq} within this approximation reduces to a set of the nonlinear equations,
\begin{align}\label{sPmodel}
\frac{d\rho_m^{ge}}{dt}&=(i\delta-\gamma)\rho_m^{ge}\nonumber\\
&\phantom{=}\,+i(1-2\rho_m^{ee})\left[\mathcal{R}\psi(\mathbf{r}_m)+\sum_{k\ne m}\mathcal{H}_{mk}\rho_k^{ge}\right],\nonumber\\
\frac{d\rho_m^{ee}}{dt}&=-2\gamma \rho_m^{ee}\nonumber\\
&\phantom{=}\,-2\operatorname{Im}\left[\rho_m^{eg}\mathcal{R}\psi(\mathbf{r}_m)+\rho_m^{eg}\sum_{k\ne m}\mathcal{H}_{mk}\rho_k^{ge}\right],
\end{align}
where $\rho_m^{ee}(t)=\left<\sigma_m^{ee}(t)\right>$ and $\rho_m^{eg}=(\rho_m^{ge})^*$ [$\rho_m^{ee}(t)+\rho_m^{gg}(t)=1$]. 
In the absence of light-established inter-atomic coupling terms, i.e.\ the terms proportional to $\mathcal{H}_{mk}$, the model becomes equal to independent-atom optical Bloch equations. 
Equations \eqref{sPmodel} can be numerically solved to obtain a semiclassical prediction $n_C^\text{SC}$ for the steady-state coherent scattering rate.

To compare the SC model with the full quantum solution, we plot in Fig.~\ref{cohSca}(a) the relative deviation $(n_C^\text{lin}-n_C^\text{SC})/n_C^\text{SC}$ for drives mode-matched and resonant with the most subradiant and superradiant collective modes. The SC model captures well the linear dependence of $(n_C^\text{lin}-n_C)/n_C$ on $I_\text{in}$, with a slope that depends strongly on the collective mode linewidth. This will be explored further in Sec.~\ref{linewidth}.

\subsection{Incoherent scattering}

The linear classical oscillator model derived for the coherent scattering from laser-driven atoms in the limit of low light intensity can only describe incoherent scattering for the atomic ensembles where the positions of atoms are spatially fluctuating. For atoms at fixed spatial positions, the only contribution to incoherent scattering are correlation functions such as $\left<\sigma_m^+\sigma_n^-\right>$ and $\left<\sigma_m^{ee}\right>$, both of which are neglected in the linear classical oscillator model.\footnote{A single photon can give rise to incoherent scattering, as is evident from considering a single isolated atom at the origin. The incoherently scattered light intensity is then $2c|G(\mathbf{r})\mathbf{d}|^2(\left<\sigma^{ee}\right>-|\left<\sigma^-\right>|^2)/\epsilon_0$. Absorption of a single photon by an atom in the ground state results in $\left<\sigma^{ee}\right>=1$ and $\left<\sigma^-\right>=0$. The likelihood of an incoherent photon emission then occurring by time $t$ after the absorption is $1-e^{-2\gamma t}$.} The presence of non-negligible incoherent scattering in our system therefore provides another signature of scattered light beyond the predictions of the linear classical oscillator model [Eq.~\eqref{lowLight}]. By gradually increasing the drive intensity we can determine at what point the incoherent scattering becomes appreciable. In Fig.~\ref{cohSca}(b) we plot the ratio of the incoherent to coherent scattering rate as a function of drive intensity, obtained from the steady-state solution of Eq.~\eqref{masterEq}, using the same parameters as Fig.~\ref{cohSca}(a). The behavior of $n_I/n_C$ is very similar to that of $(n_C^\text{lin}-n_C)/n_C$ displayed in Fig.~\ref{cohSca}(a). Indeed, we find numerically that $n_C^\text{lin}-n_C\approx 2n_I+\mathcal{O}(I_\text{in}^6)$. Hence, in this case, nonlinear coherent scattering and appreciable incoherent scattering provide an equivalent signature for the validity of the linear classical oscillator model. Note, though, that this will not hold for general detector geometries, as incoherently scattered light is distributed everywhere, whereas coherently scattered light can be very directional.

\begin{figure}
\includegraphics[trim=5cm 4cm 5cm 4cm,clip=true,width=0.48\textwidth]{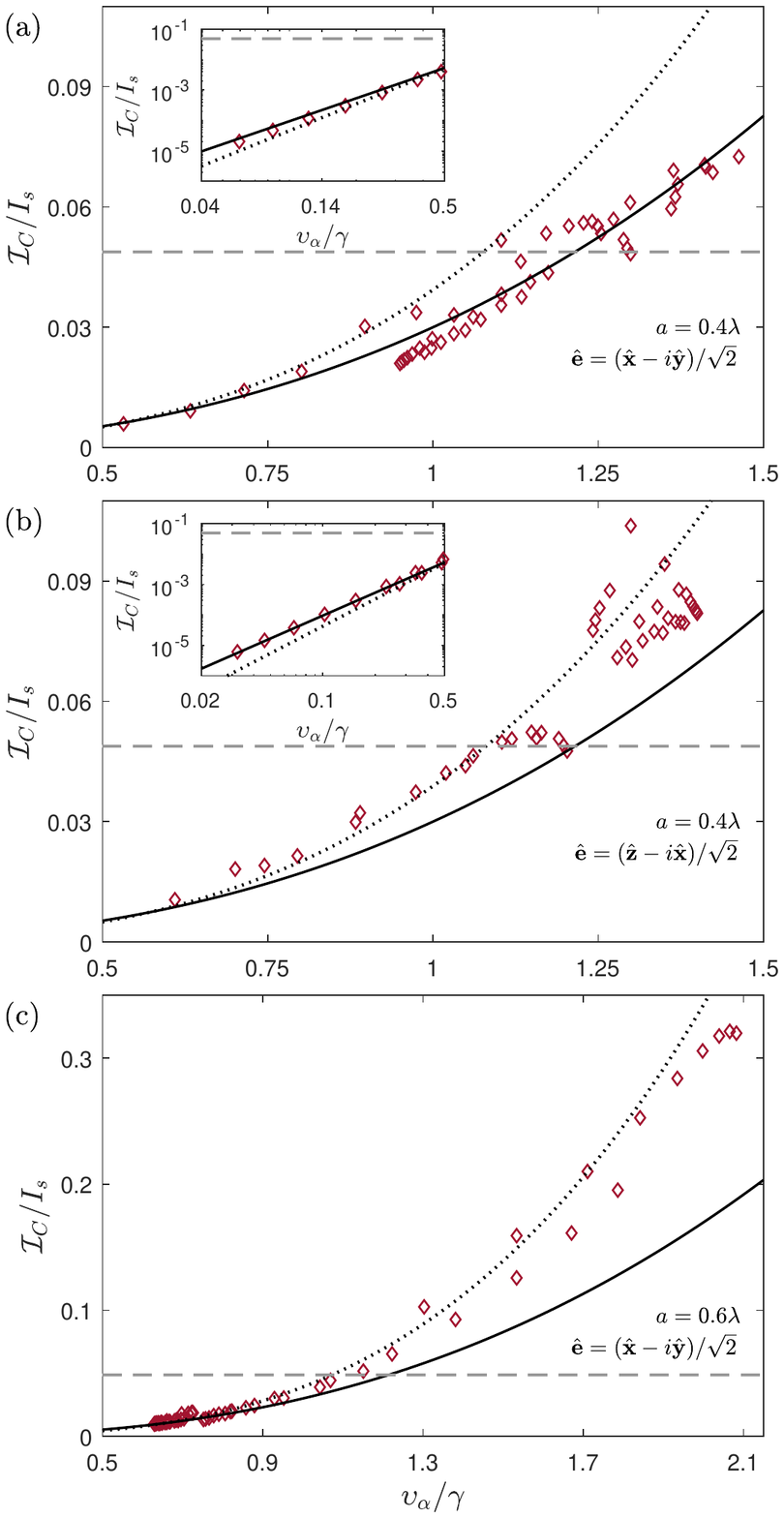}
\caption{\label{Ic}Driving the atom chain with a field that is mode-matched and resonant with a low light intensity collective eigenmode $\mathbf{u}_\alpha$ results in a critical intensity $\mathcal{I}_C$ that depends strongly on the collective mode linewidth $\upsilon_\alpha$. The atom chains used are (a) $a=0.4\lambda$, $\hat{\mathbf{e}}=(\hat{\mathbf{x}}-i\hat{\mathbf{y}})/\sqrt{2}$, (b) $a=0.4\lambda$, $\hat{\mathbf{e}}=(\hat{\mathbf{z}}-i\hat{\mathbf{x}})/\sqrt{2}$, and (c) $a=0.6\lambda$, $\hat{\mathbf{e}}=(\hat{\mathbf{x}}-i\hat{\mathbf{y}})/\sqrt{2}$. In each case we plot points for all collective modes of all atom numbers from $N=2$ to $N=10$. A power law fit $\mathcal{I}_C\propto \upsilon_\alpha^{2.5}$ (solid curves) describes well the data in (a), as well as modes $\upsilon_\alpha\lesssim 0.5\gamma$ in (b) (insets, with log-log scale). A powerlaw fit to $\mathcal{I}_C$ calculated from the semiclassical model~\eqref{sPmodel} (dotted curves) gives good agreement for drives mode-matched to collective modes with $\upsilon_\alpha\gtrsim 0.5\gamma$ in (b) and (c). In each figure, the grey dashed horizontal lines give the value of $\mathcal{I}_C$ for a single isolated atom.}
\end{figure}

\subsection{Variation with collective mode linewidth}\label{linewidth}
To quantify more precisely the behaviour of coherently scattered light  in Fig.~\ref{cohSca}(a) we introduce a critical intensity $\mathcal{I}_C$ defined as the lowest intensity $I_\text{in}$ at which $n_C$ deviates from $n_C^\text{lin}$ by 10\%, i.e., $n_C^\text{lin}-n_C=0.1n_C$. In Fig.~\ref{Ic} we plot $\mathcal{I}_C$ as a function of $\upsilon_\alpha$ for drives mode-matched and resonant with the different low light intensity collective modes of atom chains with (a) $a=0.4\lambda$, $\hat{\mathbf{e}}=(\hat{\mathbf{x}}-i\hat{\mathbf{y}})/\sqrt{2}$, (b) $a=0.4\lambda$, $\hat{\mathbf{e}}=(\hat{\mathbf{z}}-i\hat{\mathbf{x}})/\sqrt{2}$, and (c) $a=0.6\lambda$, $\hat{\mathbf{e}}=(\hat{\mathbf{x}}-i\hat{\mathbf{y}})/\sqrt{2}$. For each chain we include points for all low light intensity collective modes for all atom numbers from $N=2$ to $N=10$. The qualitative result that $\mathcal{I}_C$ on average increases with increasing $\upsilon_\alpha$ is clearly evident, as was observed in Fig.~\ref{cohSca}(a). Indeed, the data in (a) as well as modes $\upsilon_\alpha\lesssim 0.5\gamma$ in (b) follow an empirical scaling $\mathcal{I}_C\propto \upsilon_\alpha^{2.5}$. As another example, the chain with $a=0.3\lambda$, $\hat{\mathbf{e}}=(\hat{\mathbf{x}}-i\hat{\mathbf{y}})/\sqrt{2}$ (not shown) in our simulations also exhibits $\mathcal{I}_C\propto\upsilon^{2.5}$ scaling for all modes.

For comparison, we calculate $\mathcal{I}_C$ from the SC model Eq.~\eqref{sPmodel}, for the same atom chains and drive fields. A power law fit to these values, fitted to all $\upsilon_\alpha$ data points from all atom numbers, is shown by the dashed curves in Fig.~\ref{Ic} for each lattice spacing and drive polarization. The fits predict a scaling very close to $\mathcal{I}_C\propto\upsilon_\alpha^3$ for all $\upsilon_\alpha$. The chains in (b) and (c) show reasonable agreement with the SC scaling for $\upsilon_\alpha\gtrsim 0.5\gamma$. Also shown in Fig.~\ref{Ic} is the value of $\mathcal{I}_C$ for a single isolated atom, obtained from Eq.~\eqref{nSingleAtom}. This gives $\mathcal{I}_C\approx 0.05$, which lies above the values of $\mathcal{I}_C$ for many atoms for $\upsilon_\alpha\lesssim \gamma$ and below for $\upsilon_\alpha\gtrsim\gamma$.

The observation that superradiant modes follow more closely the SC model than subradiant modes is expected~\cite{bettles2019}: the dipoles of a superradiant mode are much more uniform, and hence the mean field contribution of these dominates over the quantum fluctuations. Increasing the lattice spacing from $a=0.4\lambda$ (Fig.~\ref{Ic}(a)) to $a=0.6\lambda$ (Fig.~\ref{Ic}(c)) reduces the strength of the resonance dipole-dipole interactions, which will also reduce the likelihood of quantum fluctuations, and hence give better agreement with the SC model~\cite{Kramer2015a}. For a lattice spacing of $a=0.4\lambda$, $(ka)^{-1}\approx 0.4$ and hence the dominant term in the radiation kernel~\eqref{Gdef} is the term $\propto (kr)^{-1}$. Changing the dipole orientation from $\hat{\mathbf{e}}=(\hat{\mathbf{x}}-i\hat{\mathbf{y}})/\sqrt{2}$ to $\hat{\mathbf{e}}=(\hat{\mathbf{z}}-i\hat{\mathbf{x}})/\sqrt{2}$ reduces this term  by a factor of two, hence reducing the dipole-dipole interactions, which may account for the better agreement with the SC model in Fig.~\ref{Ic}(b) compared to Fig.~\ref{Ic}(a).

We can also quantify the accuracy of the linear classical oscillator model by an intensity $\mathcal{I}_I$ at which incoherent scattering becomes appreciable. We take this to be the intensity $I_\text{in}$ at which $n_I=0.1n_C$. As in Fig.~\ref{cohSca}, we find that $\mathcal{I}_I$ follows closely the behavior of $\mathcal{I}_C$, with $\mathcal{I}_I/\mathcal{I}_C\approx 2$, independent of $\alpha$.

\begin{figure}
\includegraphics[trim=2cm 0cm 2cm 2.3cm,clip=true,width=0.48\textwidth]{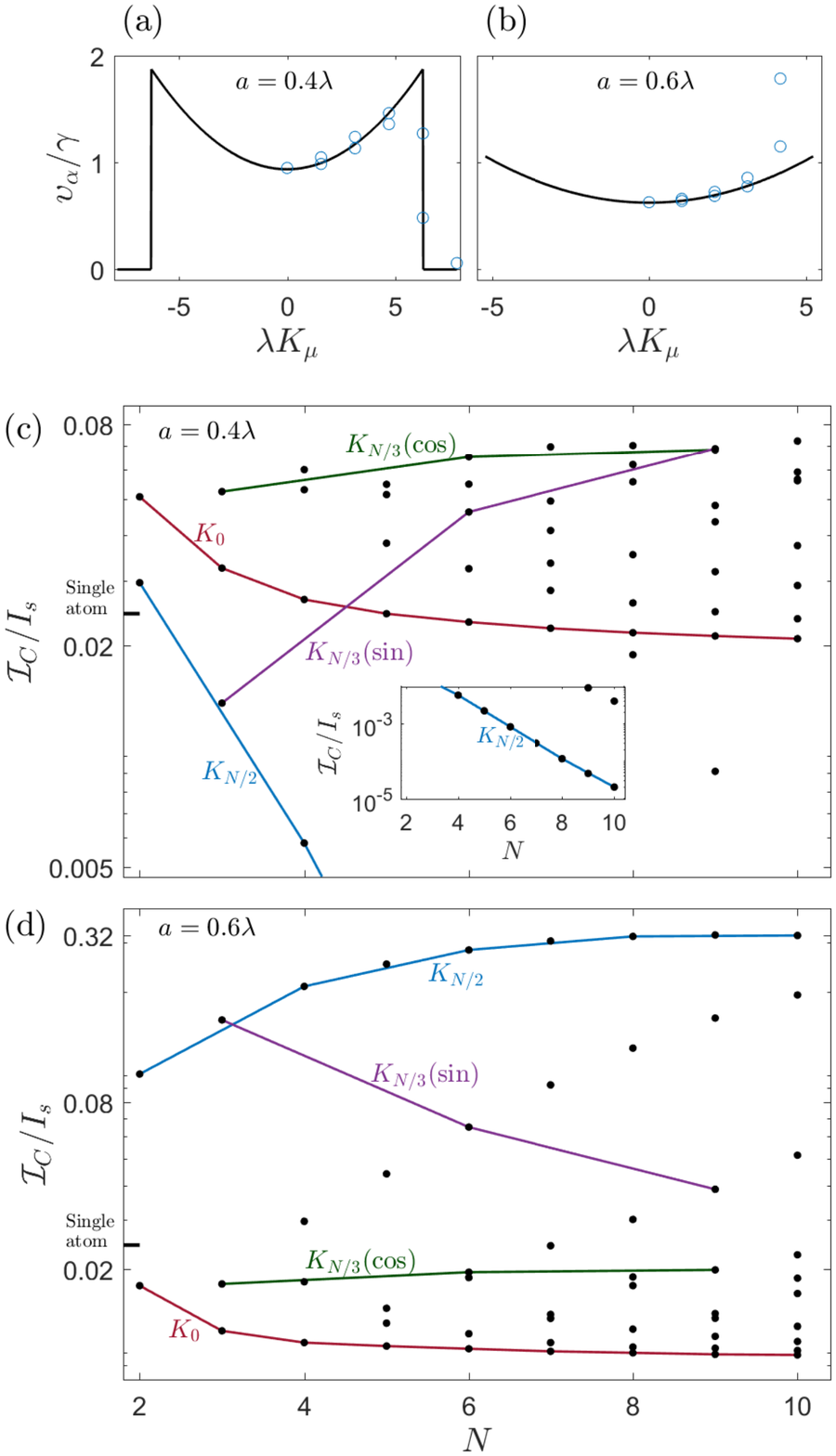}
\caption{\label{IcN}(a),(b) The collective radiative linewidths $\upsilon_\alpha$ of an infinite chain of atoms for $\hat{\mathbf{e}}=(\hat{\mathbf{x}}-i\hat{\mathbf{y}})/\sqrt{2}$ for (a) $a=0.4\lambda$ and (b) $a=0.6\lambda$. The spectrum drops abruptly to zero outside the light line, $K_\mu>2\pi/\lambda$. The linewidths of each mode of chains with 10 atom are shown for comparison (blue circles), at a wavevector (chosen to be positive) of the standing wave with maximum overlap with the mode. (c),(d) The critical intensity $\mathcal{I}_C$ as a function of atom number, for drives mode-matched and resonant with each of the $N$ low light intensity collective modes $\mathbf{u}_\alpha$ (black dots). Results for lattice spacings (c) $a=0.4\lambda$ and (d) $a=0.6\lambda$, for $\hat{\mathbf{e}}=(\hat{\mathbf{x}}-i\hat{\mathbf{y}})/\sqrt{2}$. The solid curves join the collective modes of different atom numbers that have maximum overlap with the same $K_\mu$, for $K_\mu=\pi/a$ (blue curve), $K_\mu=0$ (red curve), and $K_\mu=2\pi/(3a)$ for a cosine mode (green curve) and a sine mode (purple curve). The $K_{N/2}$ mode in (c) lies outside the light line and the corresponding $\mathcal{I}_C$ decreases exponentially with increasing $N$ (inset, with log vertical scale). The single atom value of $\mathcal{I}_C/I_s$ is indicated by a thick horizontal marker on the vertical axis.}
\end{figure}

\subsection{Variation with atom number}

For a sufficiently large atom chain, the collective eigenmodes of a chain of atoms become those of the Bloch waves, obtained using periodic boundary conditions,
\begin{align}\label{standingWaves}
u_{\alpha m}\xrightarrow[N\rightarrow\infty]{}A_\mu\sin (K_\mu r_m)\text{ or }B_\mu\cos(K_\mu r_m),
\end{align}
with $K_\mu=2\pi \mu/(Na)$, $\mu=0,1,...,\operatorname{floor}(N/2)$, and $A_\mu,B_\mu$ normalization factors. This provides collective linewidths in the infinite chain limit and 
a reasonable approximation for the eigenmodes of many finite systems also~\cite{Asenjo_prx}. We plot these in Fig.~\ref{IcN} for $\hat{\mathbf{e}}=(\hat{\mathbf{x}}-i\hat{\mathbf{y}})/\sqrt{2}$ with lattice spacing (a) $a=0.4\lambda$ and (b) $a=0.6\lambda$. For $a=0.6\lambda$, the most subradiant mode occurs at $K_\mu=0$. The $a=0.4\lambda$ chain has a similar spectrum for $K_\mu\le 2\pi/\lambda$ and then drops abruptly to zero. The light line at $K_\mu=2\pi/\lambda$ separates radiating modes from completely dark modes, and corresponds to the point where the wavevector of light radiating perpendicular to the atom chain changes from a radiating field to an evanescent field~\cite{Asenjo_prx}.

In Figs.~\ref{IcN}(c),(d) we show the critical intensity $\mathcal{I}_C$ for $\hat{\mathbf{e}}=(\hat{\mathbf{x}}-i\hat{\mathbf{y}})/\sqrt{2}$, as in Figs.~\ref{Ic}(a),(c), but now as a function of $N$. We assign Bloch waves $w^a(\mathbf{r}_m)=A_\mu\sin (K_\mu r_m)$ or $w^b(\mathbf{r}_m)=B_\mu\cos(K_\mu r_m)$ to  the finite lattice collective modes $\mathbf{u}_\alpha$ by maximizing the overlap $|\sum_m w^{a,b}(\mathbf{r}_m)u_{\alpha m}|$. We can then join points from different atom numbers that have the same $K_\mu$ and standing wave parity, e.g,, $K_0=0$ (for all $N$), $K_{N/2}=\pi/a$ (for even $N$), and $K_{N/3}=2\pi/(3a)$ (for $N$ that is a multiple of 3). For a chain with $a=0.4\lambda$, the $K_\mu=\pi/a$ curve decreases exponentially with increasing $N$ [inset to Fig.~\ref{IcN}(c)]. This wavevector resides outside the light line and therefore gives a linewidth that goes to zero in the limit of an infinite chain. Note also the broken degeneracy of the $K_{N/3}$ curves, due to finite size effects. The separation between the solutions decreases as $N$ increases, and in the infinite chain limit these two curves will coincide. The collective linewidths of the chains with 10 atoms are shown in Fig.~\ref{IcN}(a),(b), at a wavevector (chosen to be positive) of the standing wave that has maximimum overlap with the given collective mode. These follow closely the infinite chain result.

\section{Standing wave driving fields}\label{realisticDrives}
\begin{figure}
\includegraphics[trim=1cm 9cm 1cm 8cm,clip=true,width=0.48\textwidth]{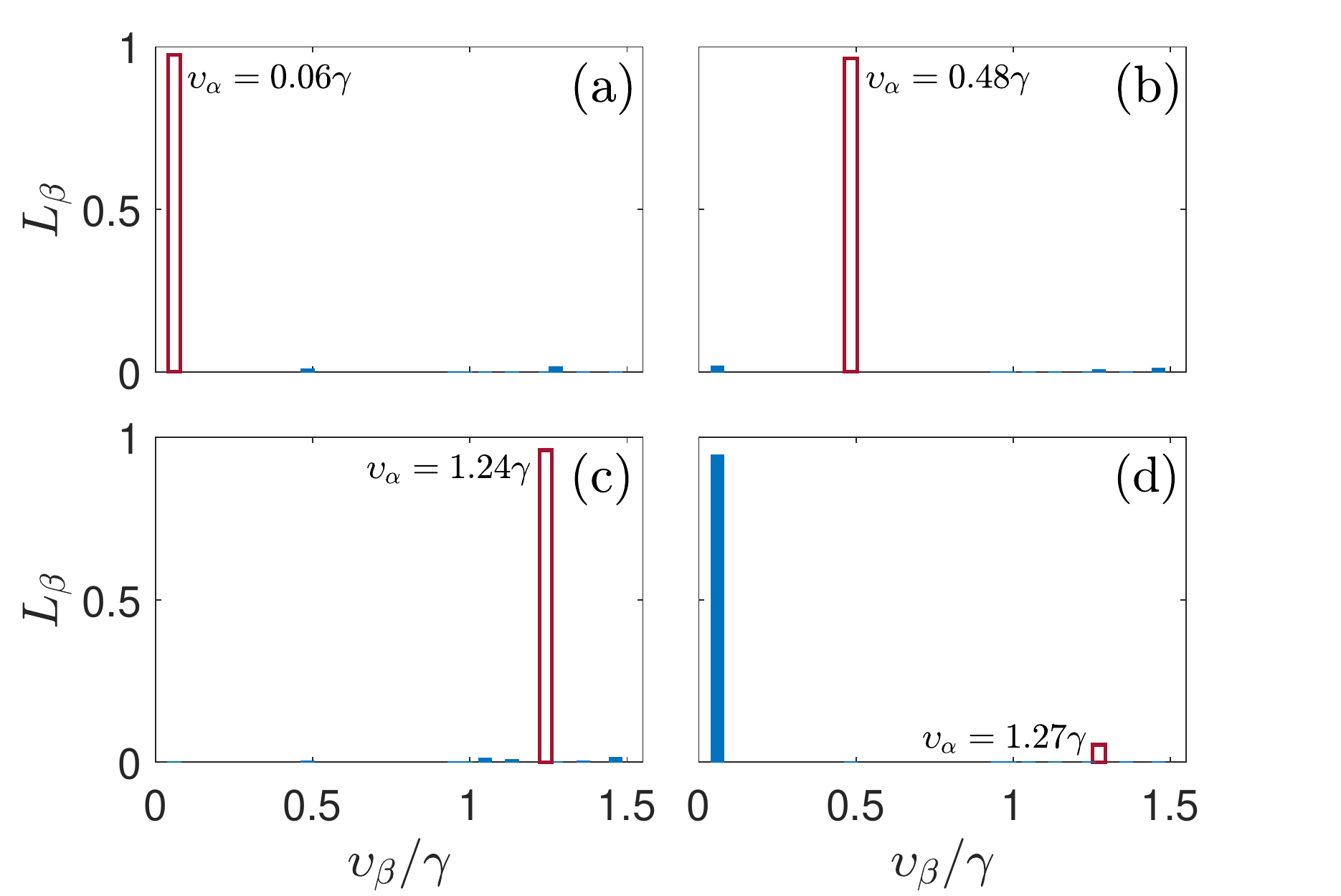}
\caption{\label{overlap} Occupation measure $L_\beta$ of each of the collective modes $\beta$ with the steady-state polarization obtained in the limit of low light intensity, for a standing-wave drive targeting mode $\alpha$. The linewidth of the targeted modes are indicated in the figures, with occupations given by the unfilled red bar. The filled blue bars give the occupations of the remaining modes. In cases (a)-(c) the overlap is dominated by the targeted mode, even in (a) where the targeted mode lies outside the light line. An anomaly occurs in (d) where the most subradiant mode dominates the occupation despite the drive targeting a superradiant mode. This is due to a Fano resonance between the two modes, as discussed further in the main text.}
\end{figure}

\begin{figure}
\includegraphics[trim=3cm 0.5cm 3cm 0.5cm,clip=true,width=0.48\textwidth]{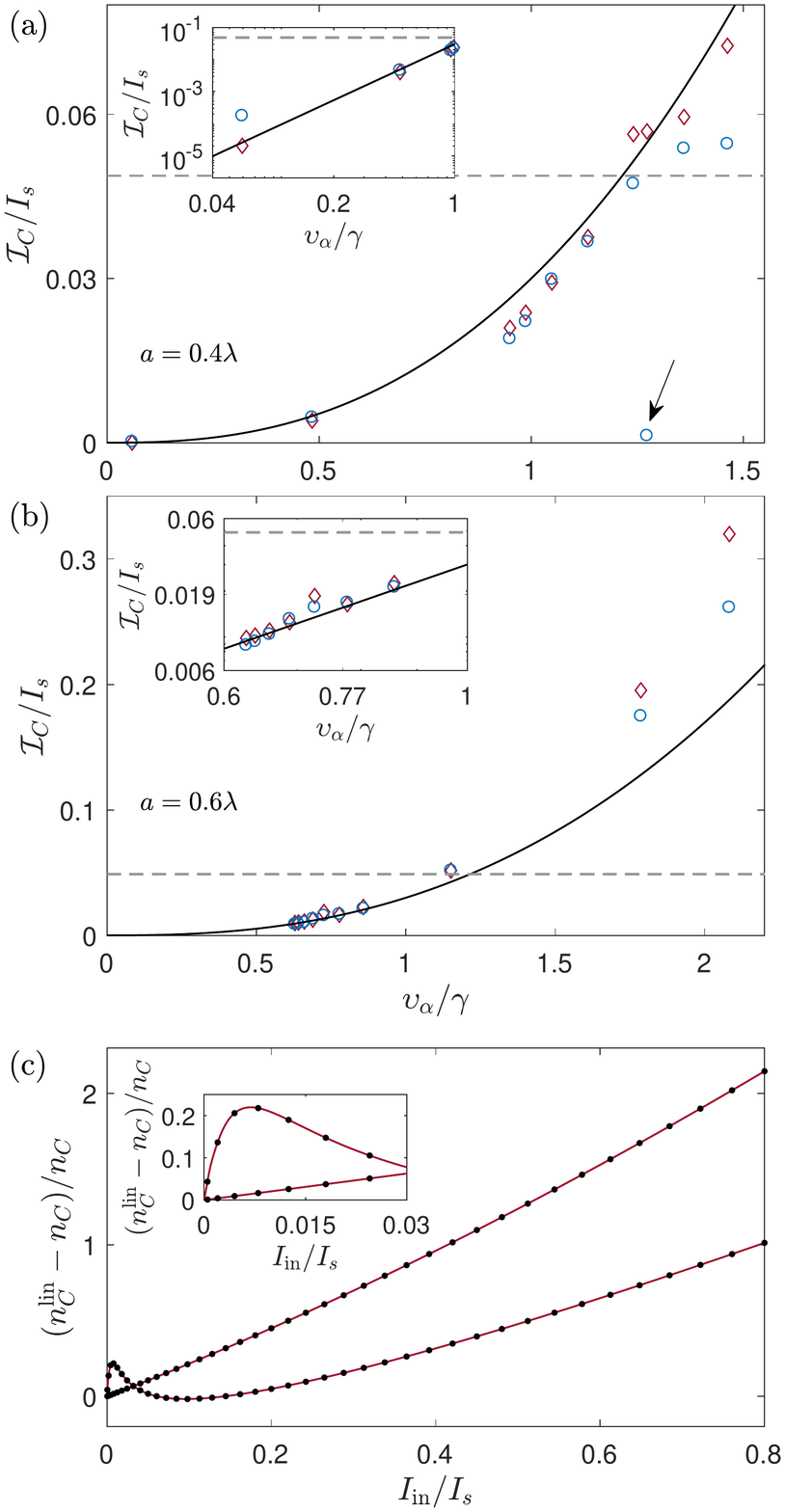}
\caption{\label{Ireal}(a),(b) The critical intensity $\mathcal{I}_C$ for both the optimized standing wave drives (blue circles) and the perfectly mode-matched drives (red diamonds), as a function of collective mode linewidth of the targeted mode. The optimized standing wave drive agrees well for the perfectly mode-matched drive for most cases. Solid curves show the $\upsilon_\alpha^{2.5}$ scaling from Fig.~\ref{Ic}. (a) $a=0.4\lambda$. The most subradiant collective mode lies outside the light line, for which the standing wave drive gives a substantially higher $\mathcal{I}_C$ than the perfectly mode-matched drive. At $\upsilon_\alpha=1.27\gamma$ the standing wave drive gives an anomalously low $\mathcal{I}_C$ (point indicated by an arrow). (b) $a=0.6\lambda$. All the collective modes lie inside the light line, and the standing wave drive and perfectly mode-matched drive give comparable $\mathcal{I}_C$ for all $\upsilon_\alpha$. (c) Deviation between the full quantum scattering and scattering predicted by the linear classical oscillator model for the $a=0.4\lambda$ atom chain driven by a standing wave drive overlapping with the $\upsilon_\alpha=1.24\gamma$ mode (monotonic curve) and the $\upsilon_\alpha=1.27\gamma$ mode (non-monotonic curve) (dots are numerical simulations, red curves are spline fits). A Fano resonance at $\upsilon_\alpha=1.27\gamma$ results in an anomalously high $\mathcal{I}_C$.}
\end{figure}

Next we drive the atoms with fields $\psi^a(\mathbf{r}_m)=A_\mathbf{k}\sin(\mathbf{k}\cdot\mathbf{r}_m)$ or $\psi^b(\mathbf{r}_m)=B_\mathbf{k}\cos(\mathbf{k}\cdot\mathbf{r}_m)$, with $\mathbf{k}$ the wavevector of the incident light and $A_\mathbf{k},B_\mathbf{k}$ normalization factors, that coincide with the Bloch waves $w^{a,b}(K_\mu r_m)$ [Eq.~\eqref{standingWaves}] with wavenumber $K_\mu=\mathbf{k}\cdot\hat{\mathbf{z}}=k\cos\theta$. Here $\theta$ is the angle between $\mathbf{k}$ and the atom chain. By changing $\theta$ we can therefore target modes with different $K_\mu\le 2\pi/\lambda$. Modes with $K_\mu>2\pi/\lambda$ lie outside the light line in Fig.~\ref{IcN}(a) and, in the infinite lattice limit, cannot be excited due to their rapid phase variation. For a finite chain we can find an optimal standing wave drive $\psi(\mathbf{r}_m)$ by maximizing $|\sum_m \psi^{a,b}(\mathbf{r}_m)u_{\alpha m}|$ for a given $\alpha$ and varying $\psi^{a,b}$. The targeting is further enhanced by tuning the drive frequency to the resonance of the targeted mode, such that $\zeta_\alpha=0$.

The effectiveness of the standing waves to target a particular eigenstate can be quantified by the occupation measure $L_\beta$ of the exact eigenstates in the steady-state polarization, defined by~\cite{Facchinetti16}
\begin{align}\label{Ldef}
L_\beta=\frac{\left|\sum_m u_{\beta m}\left<\sigma_m^-\right>\right|^2}{\sum_\eta\left|\sum_m u_{\eta m}\left<\sigma_m^-\right>\right|^2}
\end{align}
with$\left<\sigma_m^-\right>$ calculated using the low light intensity equation~\eqref{lowLight}. [Note the absence of a complex conjugation of the $u_{\alpha m}$. The modes $\mathbf{u}_\alpha$ are not orthogonal, but they do satisfy the biorthogonality condition $\mathbf{u}_\alpha^T\mathbf{u}_\beta=\delta_{\alpha\beta}$ except for possible (rare) cases when $\mathbf{u}_\alpha^T\mathbf{u}_\alpha=0$. Hence a transpose rather than a conjugate transpose is used in Eq.~\eqref{Ldef}.] Example distributions of $L_\beta$ are shown in Fig.~\ref{overlap} for an atom chain with $a=0.4\lambda$, $\hat{\mathbf{e}}=(\hat{\mathbf{x}}-i\hat{\mathbf{y}})/\sqrt{2}$, for drive fields targeting four different low light intensity collective modes. In all but one case, the occupation is dominated by the targeted mode, even for the mode with $\upsilon_\alpha=0.06\gamma$, which resides outside the light line. The anomalous case Fig.~\ref{overlap}(d) is due to a Fano resonance between the targeted mode and the most subradiant mode, and will be discussed further shortly.

Using standing light waves, we can carry out an analogous study to that in Fig.~\ref{Ic}. In Fig.~\ref{Ireal} we plot the resulting $\mathcal{I}_C$ as a function of the collective linewidths $\upsilon_\alpha$ of the targeted mode for chains of 10 atoms for (a) $a=0.4\lambda$ and (b) $a=0.6\lambda$, with $\hat{\mathbf{e}}=(\hat{\mathbf{x}}-i\hat{\mathbf{y}})/\sqrt{2}$. Also included are the results for $\mathcal{I}_C$ using a perfectly mode-matched drive [from Fig.~\ref{Ic}(a),(c)]. Both the standing-wave drive and the perfectly mode-matched drive give comparable $\mathcal{I}_C$ for the $a=0.6\lambda$ atom chain. For the $a=0.4\lambda$ chain, the standing wave and perfectly mode-matched drive give comparable $\mathcal{I}_C$ for targeted modes with $\upsilon_\alpha\gtrsim 0.5\gamma$, aside from one anomalous point at $\upsilon_\alpha=1.24\gamma$ that will be discussed shortly. The standing-wave drive gives a substantially larger $\mathcal{I}_C$ for the most subradiant mode in Fig.~\ref{Ireal}(a), which lies outside the light line and hence overlap between the standing wave and this collective mode is small. We find similar results for a chain $10$ of atoms with $a=0.3\lambda$ (not shown), with the two subradiant modes residing outside the light line giving substantially larger $\mathcal{I}_C$ for a standing-wave drive compared to a perfectly mode-matched drive.

We now explore further the anomalous point in Fig.~\ref{Ireal}(a) at $\upsilon_\alpha=1.27\gamma$. In Fig.~\ref{Ireal}(c) we plot the relative deviation $(n_C^\text{lin}-n_C)/n_C$, where the drive $\psi_{1.27}(\mathbf{r})$ is a standing wave overlapping with the $\upsilon_\alpha=1.27\gamma$ mode. The result for a standing wave drive overlapping with the $\upsilon_\alpha=1.24\gamma$ mode is also shown for comparison. The initial growth for the $\psi_{1.27}(\mathbf{r})$ drive is much more rapid than the linewidth $\upsilon_\alpha=1.27\gamma$ alone would suggest, resulting in a much lower $\mathcal{I}_C$, close to that of the most subradiant modes. The occupation measures $L_\beta$ for these two drives are shown in Fig.~\ref{overlap}(c),(d). The $\psi_{1.24}(\mathbf{r})$ drive [(c)] leads to a predominant occupation in the targeted mode, whereas the $\psi_{1.27}(\mathbf{r})$ drive [(d)] predominantly targets the most subradiant mode. This is due to the nonorthogonality between the superradiant mode with $\upsilon_\alpha=1.27\gamma$ and the most subradiant mode with $\upsilon_\beta=0.059\gamma$ is $|\mathbf{u}_\alpha^\dagger\mathbf{u}_\beta|\approx 0.3$, which is appreciable. Furthermore, the collective level shift of the most subradiant mode lies within the linewidth of the $\upsilon_\alpha=1.27$ mode (both collective level shifts are $\approx 1.0\gamma$). The combination of these effects leads to a Fano resonance between the $\upsilon_\alpha=1.27\gamma$ mode and the $\upsilon_\beta=0.059\gamma$ mode, resulting in a suppressed scattering rate~\cite{Facchinetti16,ruostekoski2017}. This is likely responsible for the  much lower $\mathcal{I}_C$.

\section{Conclusions}\label{conclusion}

We compared light scattering obtained from the linear classical oscillator model with a full quantum treatment, as a function of increasing light intensity. We showed that deviations between the two approaches become appreciable at an intensity that is much lower for drive fields targeting subradiant modes than superradiant modes, and identify scaling relationships between this intensity and the linewidth of the mode being driven. The scaling relationships are largely insensitive to atom number, lattice orientation, lattice spacing and precise drive field profile.  An SC model captures the qualitative conclusions of our results and, for superradiant modes, many of the quantitative features also. 
It would be interesting to test how well the results carry over to higher dimensional lattices and lattices with different geometries.
Further work could also explore the effects of fluctuating positions~\cite{Jenkins2012a} on our findings, in which case incoherent scattering can also occur within the linear classical oscillator model. For example, it would be interesting to compare the importance of incoherent scattering arising from position fluctuations with incoherent scattering arising from excited state occupations.

\section{Acknowledgements}

We acknowledge financial support from EPSRC.

\appendix*
\section{Evaluation of the far-field interference integral}\label{appen}
Substitution of Eq.~\eqref{Escat} into Eqs.~\eqref{scatteringrate1}--\eqref{scatteringratei} gives the interference integrals
\begin{align}\label{intG}
I_{mn}=\frac{2c}{\hbar\epsilon_0\omega}\int_S dS \left[\mathsf{G}(\mathbf{r}-\mathbf{r}_m)\mathbf{d}\right]^*\mathsf{G}(\mathbf{r}-\mathbf{r}_n)\mathbf{d}.
\end{align}
For a detector sufficiently far from the atoms we can take the far-field (Fraunhofer) limit for the scattering kernel $\mathsf{G}$,
\begin{align}
\mathsf{G}(\mathbf{r}-\mathbf{r}_m)\mathbf{d}\xrightarrow[r\rightarrow\infty]{}\frac{k^2}{4\pi r}e^{ikr}e^{-ik\hat{\mathbf{r}}\cdot\mathbf{r}_m}(\hat{\mathbf{r}}\times \mathbf{d})\times \hat{\mathbf{r}}.
\end{align}
Using $dS=r^2d\theta d\phi \sin\theta$, for polar angle $\theta$ and azimuthal angle $\phi$, this gives
\begin{align}\label{Imnint}
I_{mn}=\frac{3\gamma}{4\pi}\int_S d\theta d\phi\,\sin\theta\left(1-|\hat{\mathbf{r}}\cdot\hat{\mathbf{e}}|^2\right)e^{ik\hat{\mathbf{r}}\cdot\mathbf{r}_{mn}}.
\end{align}
In our choice of coordinate system we have $\hat{\mathbf{r}}_{mn}=\hat{\mathbf{z}}$ ($\hat{\mathbf{r}}_{mn}=\mathbf{r}_{mn}/|\mathbf{r}_{mn}|$). For a detector that completely encloses the atoms, the integral is over a full $4\pi$ surface and hence
\begin{widetext}
\begin{align}
I_{mn}&=\frac{3\gamma}{4\pi}\int_0^\pi d\theta \int_0^{2\pi}d\phi\,\sin\theta (1-|\hat{d}_x\cos\phi\sin\theta+\hat{d}_y\sin\phi\sin\theta+\hat{d}_z\cos\theta|^2)e^{ikr_{mn}\cos\theta}\nonumber\\
&=\frac{3\gamma}{4}\int_0^\pi d\theta \,\sin\theta (2-2|\hat{d}_z|^2-(1-3|\hat{d}_z|^2)\sin^2\theta)e^{ikr_{mn}\cos\theta}\nonumber\\
&=3\gamma(1-|\hat{d}_z|^2)\frac{\sin kr_{mn}}{kr_{mn}}+3\gamma(1-3|\hat{d}_z|^2)\left(\frac{\cos kr_{mn}}{k^2r_{mn}^2}-\frac{\sin kr_{mn}}{k^3r_{mn}^3}\right)\nonumber\\
&=3\gamma(1-|\hat{\mathbf{r}}_{mn}\cdot\hat{\mathbf{e}}|^2)\frac{\sin kr_{mn}}{kr_{mn}}+3\gamma(1-3|\hat{\mathbf{r}}_{mn}\cdot\hat{\mathbf{e}}|^2)\left(\frac{\cos kr_{mn}}{k^2 r_{mn}^2}-\frac{\sin kr_{mn}}{k^3r_{mn}^3}\right)\nonumber\\
&=2\gamma_{mn}.
\end{align}
\end{widetext}
In the last line we have removed reference to a coordinate system by replacing $\hat{d}_z$ by $\hat{\mathbf{r}}_{mn}\cdot\hat{\mathbf{e}}$.


%

\end{document}